\documentclass[a4paper,fleqn,usenatbib]{mnras}
\usepackage[multidot]{grffile}

\usepackage{mathptmx}

\usepackage[T1]{fontenc}
\usepackage{ae,aecompl}

\usepackage{graphicx}	
\usepackage{amsmath}	
\usepackage{amssymb}	

\usepackage{wasysym}
\usepackage{courier}
\usepackage{mathrsfs}
\usepackage{threeparttable}
\usepackage{verbatim}
\usepackage{comment}
\usepackage[british]{babel}
\usepackage{color}

\title[KIC 7107778]{Asteroseismology of KIC 7107778: a binary comprising almost identical subgiants}

\author[ ]{
Yaguang Li,$^{1,2}$\thanks{E-mail: hnwilliam@hotmail.com}
Timothy R. Bedding,$^{2,3}$\thanks{tim.bedding@sydney.edu.au}
Tanda Li,$^{2,3}$
Shaolan Bi,$^{1}$
\newauthor
{  }Simon J. Murphy,$^{2,3}$
Enrico Corsaro,$^{4}$
Li Chen$^{1}$
and Zhijia Tian$^{5}$
\\
$^{1}$Department of Astronomy, Beijing Normal University, Beijng 100875, China\\
$^{2}$Sydney Institute for Astronomy (SIfA), School of Physics, University of Sydney, NSW 2006, Australia\\
$^{3}$Stellar Astrophysics Centre, Department of Physics and Astronomy, Aarhus University, Ny Munkegade 120, DK-8000 Aarhus C, Denmark\\
$^{4}$INAF - Osservatorio Astrofisico di Catania, Via S. Sofia 78, I-95123 Catania, Italy\\
$^{5}$Department of Astronomy, Peking University, Beijing 100871, China\\
}

\date{Accepted XXX. Received YYY; in original form ZZZ}

\pubyear{2016}

\begin{document}
\label{firstpage}
\pagerange{\pageref{firstpage}--\pageref{lastpage}}
\maketitle

\begin{abstract}
We analyze an asteroseismic binary system: KIC 7107778, a non-eclipsing, unresolved target, with solar-like oscillations in both components. We used \emph{Kepler} short cadence time series spanning nearly two years to obtain the power spectrum. Oscillation mode parameters were determined using Bayesian inference and a nested sampling Monte Carlo algorithm with the \texttt{DIAMONDS} package. The power profiles of the two components fully overlap, indicating their close similarity. We modelled the two stars with \texttt{MESA} and calculated oscillation frequencies with \texttt{GYRE}. Stellar fundamental parameters (mass, radius and age) were estimated by grid modelling with atmospheric parameters and the oscillation frequencies of $l=0,2$ modes as constraints. Most $l=1$ mixed modes were identified with models searched using a bisection method. Stellar parameters for the two sub-giant stars are: $M_A=1.42\pm0.06$ $M_{\astrosun}$, $M_B=1.39\pm0.03$ $M_{\astrosun}$, $R_A=2.93\pm0.05$ $R_{\astrosun}$, $R_B=2.76\pm0.04$ $R_{\astrosun}$, $t_A=3.32\pm0.54$ Gyr and $t_B=3.51\pm0.33$ Gyr. The mass difference of the system is $\sim$1\%. The results confirm their simultaneous birth and evolution, as is expected from binary formation. KIC 7107778 comprises almost identical twins, and is the first asteroseismic sub-giant binary to be detected.
\end{abstract}




\section{Introduction}
\label{sec:intro}

Binary star systems provide an ideal astronomical laboratory to study stellar structure and evolution. The fact that two components share same metal abundance and age provides powerful constraints on models. Eclipsing binaries are especially useful, since mass and radius can be directly measured from orbits and eclipses through spectroscopic observations and precise light curves analysis \citep{1991A&ARv...3...91A}.

Asteroseismology ushers in a new way to study binaries \citep{2014arXiv1404.7501H}, even for unresolved ones \citep{2014ApJ...784L...3M}, by analyzing the two components separately. Determining stellar mass and radius is feasible for single stars with solar-like oscillations (e.g. \citealt{2010A&A...522A...1K}; \citealt{2014ApJS..210....1C}). Sub-giant and red-giant oscillating stars displaying p (pressure dominated in the envelope) and g (gravity dominated in the core) mixed modes \citep{2013ARA&A..51..353C}, are good indicators of evolutionary stages. Their great sensitivity to mass and age can produce precise estimates of stellar parameters (e.g. \citealt{2010ApJ...723.1583M}; \citealt{2012ApJ...745L..33B}). Binary systems with pulsators can be analyzed from modulation of the pulsation frequencies (\citealt{2014MNRAS.441.2515M}; \citealt{2015MNRAS.450.4475M}; \citealt{2012MNRAS.422..738S}), so that system parameters can be determined with radial velocity curves simply derived through photometry.

\cite{2016arXiv160909581W} pointed out that, until now, only five binary star systems have been detected with solar-like oscillations from both components. They are $\alpha$ Cen A and B (\citealt{1999MNRAS.303..579K}; \citealt{1998IAUS..185..285B}; \citealt{2001A&A...374L...5B}; \citealt{2003A&A...406L..23C}), 16 Cyg A and B (KIC 12069424 and KIC 12069449; \citealt{2012ApJ...748L..10M}; \citealt{2015ApJ...811L..37M}), KIC 9139163 and KIC 9139151 (\citealt{2014A&A...566A..20A}; \citealt{2012A&A...543A..54A}), HD 177412 (KIC 7510397; \citealt{2015A&A...582A..25A}), and HD 176465 (KIC 10124866, also known as Luke \& Leia; \citealt{2016arXiv160909581W}). The latter two systems are not resolved by \emph{Kepler}, such that their light variations are mixed in a single time series. Both systems were analyzed with one power spectrum, from which two sets of oscillation profiles were measured. Specifically, HD 177412 contains two main-sequence stars with mass difference $\sim$ 7.5\%, with separable oscillation frequency ranges, and HD 176465 contains two extremely similar main-sequence stars with mass difference $\sim$ 3\%, leading to two significantly overlapping oscillation ranges. In addition, KIC 10080943 is another unresolved binary system, comprising two $\delta$ Sct/$\gamma$ Dor hybrid pulsators on the main-sequence \citep{2015MNRAS.454.1792K,2015A&A...584A..35S}.

In this context, these binaries are not strictly 'twins' in that the masses are not equal to within 2 percent \citep{lucy2006,simon&obbie2009}. Twins are almost always found at small separations \citep[$P\lesssim25$\,d,][]{lucy&ricco1979}, though not all close binaries are twins. Further, they are more common among lower-mass systems, both in observational data (e.g. \citealt{tokovinin2000,simon&obbie2009}) and in binary star-formation simulations (e.g. \citealt{bate2009}). Their importance lies in their ability to discriminate between dominant physical processes operating in pre-main-sequence binaries, as those authors have discussed.

Here we report on the widely-separated twin binary system KIC 7107778, with mass difference $\sim$ 1\% based on our findings, an analog to Luke \& Leia \citep{2016arXiv160909581W}. The system was not resolved by \emph{Kepler} but was detected with oscillations from two stars in the mixed time series. It proves to be the first asteroseismic sub-giant binary system to be detected, with completely overlapping power spectra of two component stars. The paper will be structured as follows. Section~\ref{sec:obs} provides observations from previous literature and describes the processing of \emph{Kepler} data. Section~\ref{sec:modepara} illustrates the oscillation mode parameters. Section~\ref{sec:model} details our asteroseismic analysis with stellar models of two stars, and is followed by discussions and conclusions in Section~\ref{sec:con}.


\section{Observation and Data Processing}
\label{sec:obs}

KIC 7107778 is a binary system observed as a single target, with \emph{Kepler} magnitude $K_{p}=11.38$. The first realization of its binary identity was from speckle interferometry \citep{2012AJ....144..165H}. The angular separation of the two stars is $\rho=0.0288\arcsec$, measured with filters whose central wavelengths are 692nm and 880nm. The physical separation can be determined by combining luminosities from models with the apparent magnitude; however, such estimation is only approximate because the apparent magnitude is a blended contribution from both stars. We took advantage of the parallax determined from Gaia \citep{2016arXiv160904172G}, $1.74\pm0.43$ mas, to infer the distance to Earth and physical projected separation between the two stars, which are $d=574^{+189}_{-113}$ pc, and $s=16.52^{+5.44}_{-3.25}$ AU. Note that this estimation is very approxiamate because Gaia parallaxes are not yet fully reliable at the present mission stage \citep[see][]{2016A&A...595A...3F}. Assuming the orbit is circular, and adopting Kepler's third law, it suggests the orbital period of this system is about 39 yr. Therefore, we would not expect to analyze its orbit using radial velocity curves. On the other hand, such a wide separation means fully independent evolution without tidal effects or mass transfer. In the following analysis, we treat them as two single stars that have not interacted.

Several works have measured atmospheric parameters of this system. Effective temperatures measured by SDSS and IRFM are $T_{\rm{eff}}=5129\pm82$ K and $T_{\rm{eff}}=5045\pm105$ K, respectively. The binary system was also covered by the LAMOST-\emph{Kepler} project, which used LAMOST low-resolution ($R\simeq1800$\AA) optical spectra in the waveband of $3800-9000$ \AA { }and observed \emph{Kepler}-field targets in September 2014. The LAMOST Stellar Parameter Pipeline (LSP3, \citealt{2015MNRAS.448..822X}) gives $T_{\rm{eff}}=5149\pm150$ K, and metallicity [Fe/H]$=0.11\pm0.10$ dex. \cite{2015ApJ...808..187B} used the Tillinghast Reflector Echelle Spectrograph (TRES) to obtain medium resolution ($R\simeq44000$\AA) spectra in the waveband of $3800-9100$ \AA { }at Fred Lawrence Whipple Observatory. They measured the metallicity [Fe/H]$=0.05\pm0.08$ dex. However, all these observations treated two stars as one single target. We do not know to what extent the mixed value deviates from the real ones. Table~\ref{tab:atmpara} summarizes the observations in the literature.

\begin{table}
	\centering
	\caption{Atmospheric parameters of KIC 7107778.}
	\label{tab:atmpara}
	\begin{tabular}{lccc}
		\hline
Parameter	&	Value	&	Reference	\\
		\hline
$T_{\rm{eff}}$ $[\rm K]$	&	$5129\pm82$ 	 &	{\cite{2003A&A...409..205D}}	\\
				&	$5045\pm105$ 	&	{\cite{2010A&A...512A..54C}}	\\
				&	$5149\pm150$ 	&	{\cite{2015MNRAS.448..822X}}	\\
		\hline
[Fe/H] $[\rm dex]$		&	$0.11\pm0.10$ 	&	{\cite{2015MNRAS.448..822X}}	\\
				&	$0.05\pm0.08$  &	{\cite{2015ApJ...808..187B}}	\\
		\hline
	\end{tabular}
\end{table}

\begin{figure*}
	\includegraphics[width=18.7cm]{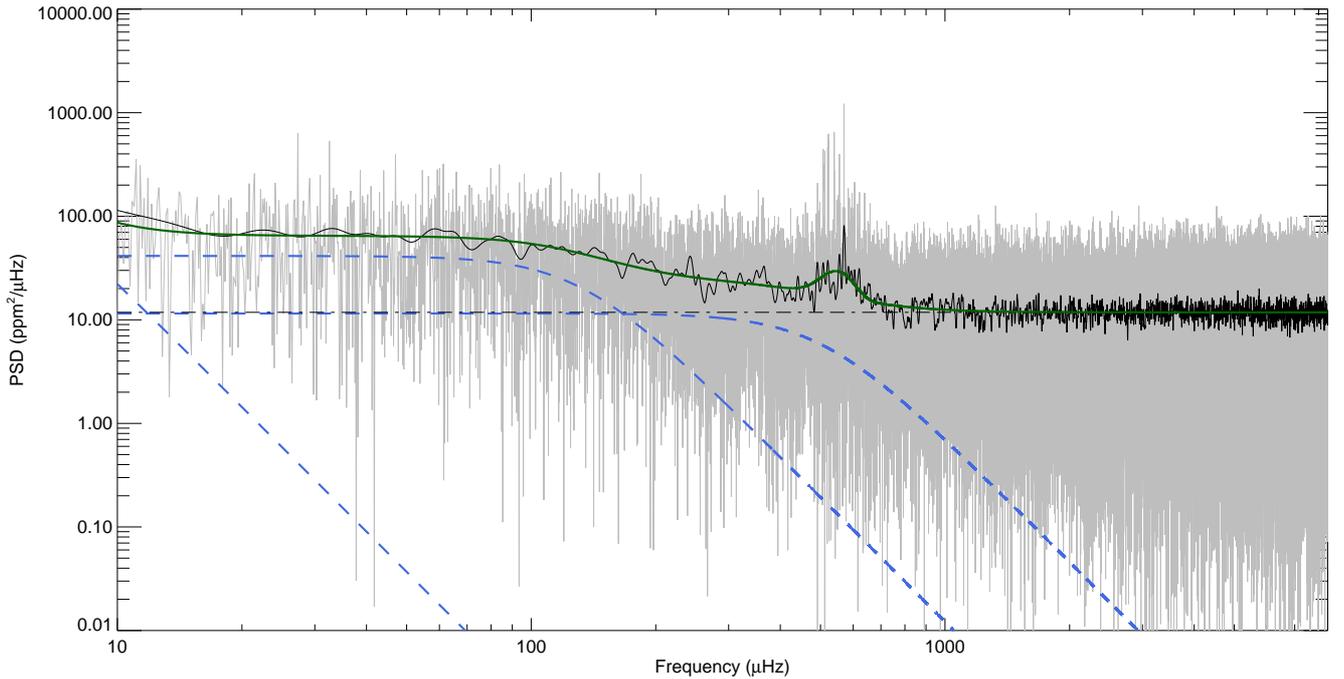}
    \caption{Power spectrum on a logarithmic scale. The solid gray line, the solid black line, the dashed blue lines, the dot-dash-dashed black line and the solid green line, outline the original power spectrum, the 6 $\mu$Hz smoothed power spectrum, the Harvey profile components, the white noise component, and the total fitted power spectrum, respectively. }
    \label{fig:ps}
\end{figure*}

\begin{figure*}
	\includegraphics[width=18.7cm]{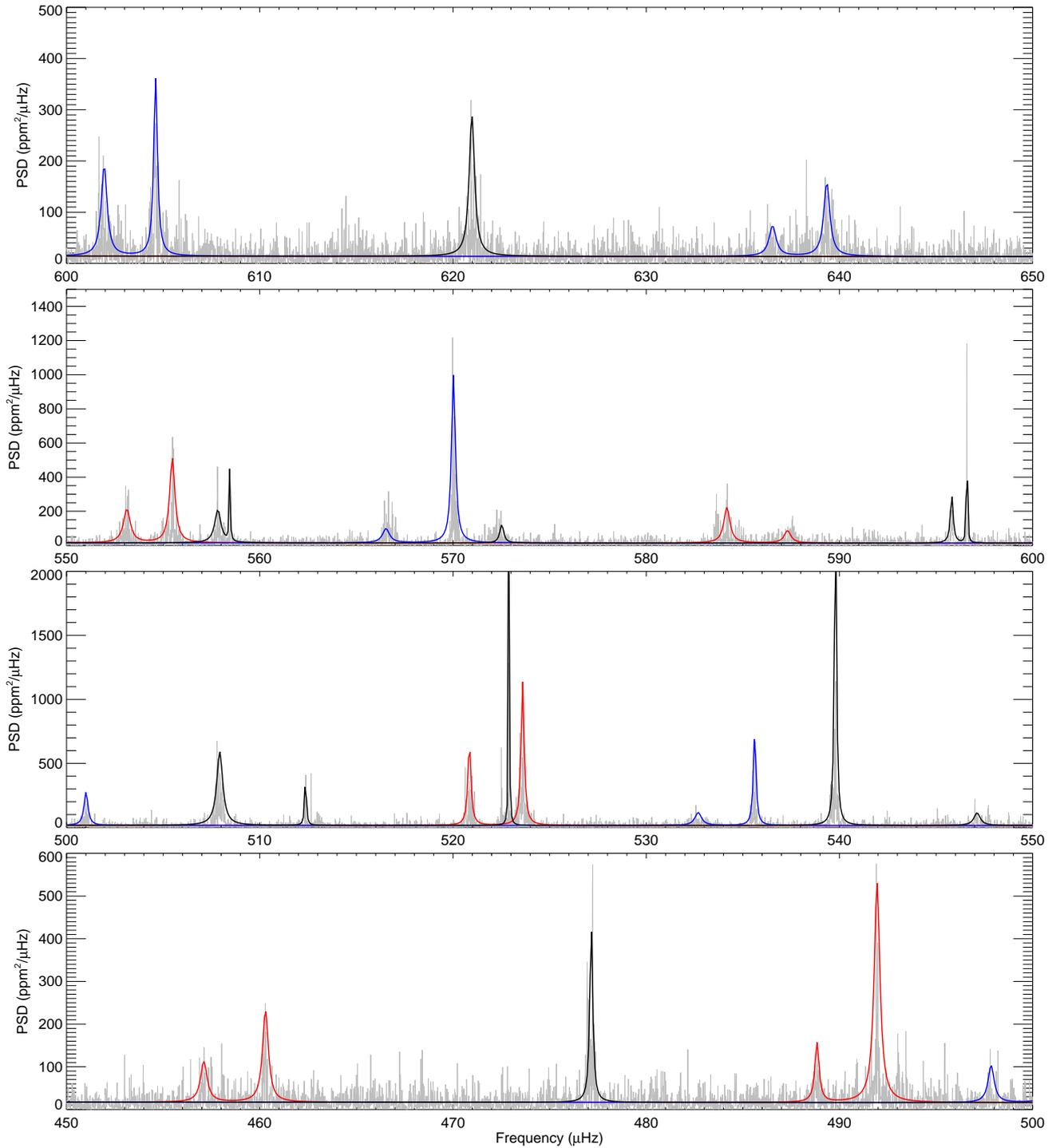}
    \caption{Power spectrum around the oscillation range on a linear scale. The gray line denotes the original power spectrum superimposed with fitted $l=1$ modes in black, $l=0$ and $l=2$ modes of star A in red, and those of star B in blue.}
    \label{fig:ps2}
\end{figure*}

The \emph{Kepler} mission observed the target in long-cadence mode (LC; 29.43 min sampling) over the whole mission and in short-cadence mode (SC; 58.84s sampling) during Q2.1, Q5 and Q7 -- Q12 (Q represents three-month-long quarters). The pulsation frequency range is centered at 550 $\mu$Hz, which is well above the Nyquist frequency of long cadence data ($\sim283$ $\mu$Hz). Therefore, we only considered the short-cadence time series. We concatenated the data and processed it following \cite{2011MNRAS.414L...6G}, correcting outliers, jumps and drifts. Then it passed through a high-pass filter which was based on a Gaussian smooth function with a width of one day. This largely minimized instrumental side-effects and only affected frequencies lower than $\sim12$ $\mu$Hz, far below the frequency range we intended to analyze. We obtained the power spectrum by applying a Lomb-Scargle Periodogram (\citealt{1976Ap&SS..39..447L}; \citealt{1982ApJ...263..835S}) to the time series with a frequency resolution $\sim 0.012$ $\mu$Hz. The power spectrum is shown in Fig.~\ref{fig:ps} and~\ref{fig:ps2} in both logarithmic and linear scales. 

The signature of solar-like oscillation is a Gaussian-like envelope located at $\nu_{\rm max}$, the so-called frequency of maximum power, and is comprised of numerous oscillation modes. In main-sequence stars, p-mode oscillations are approximately equally spaced in frequency, as described by the asymptotic equation with radial orders $n$ and spherical degrees $l$ \citep{1980ApJS...43..469T}:
\begin{equation}
\nu_{nl}=\Delta\nu(n+\frac{l}{2}+\epsilon)-\delta\nu_{0l},
\end{equation}
where $\Delta\nu$ is the large separation, which measures the spacing of adjacent modes with the same $l$, $\delta\nu_{0l}$ is the small separation, and $\epsilon$ is an offset. In more evolved stars, the core will have g-mode oscillations, which are approximately equally spaced in period. The analogous asymptotic equation is specified by order $n_g$:
\begin{equation}
\Pi_{nl}=\nu_{nl}^{-1}=\Delta\Pi_{l}(n_g+\epsilon_g),
\end{equation}
where $\Delta\Pi_{l}$ is the period spacing, and $\epsilon_g$ is an offset.

As the star evolves off the main-sequence, central hydrogen depletes, and p and g mixed modes of $l\geq1$ appear to have ``avoided crossings'' \citep{1977A&A....58...41A}, whereby oscillation frequencies are no longer equally spaced in either frequency or period. The $l=0$ p modes are unaffected, which assisted us in determining the mean large separation $\langle\Delta\nu\rangle$ in an \'{e}chelle diagram. The best value for $\langle\Delta\nu\rangle$ is the one that makes the $l=0$ ridge vertical. Fig.~\ref{fig:echelle} displays \'{e}chelle diagrams for the two stars. To avoid ambiguity, the star with smaller $\langle\Delta\nu\rangle$ is named as KIC 7107778 A, and the larger one KIC 7107778 B. The best values are: $\langle\Delta\nu\rangle_A=31.83\pm0.02$ $\mu$Hz and $\langle\Delta\nu\rangle_B=34.55\pm0.01$ $\mu$Hz.




\section{Mode Parameters}
\label{sec:modepara}

\begin{figure*}
	\includegraphics[width=7.0cm]{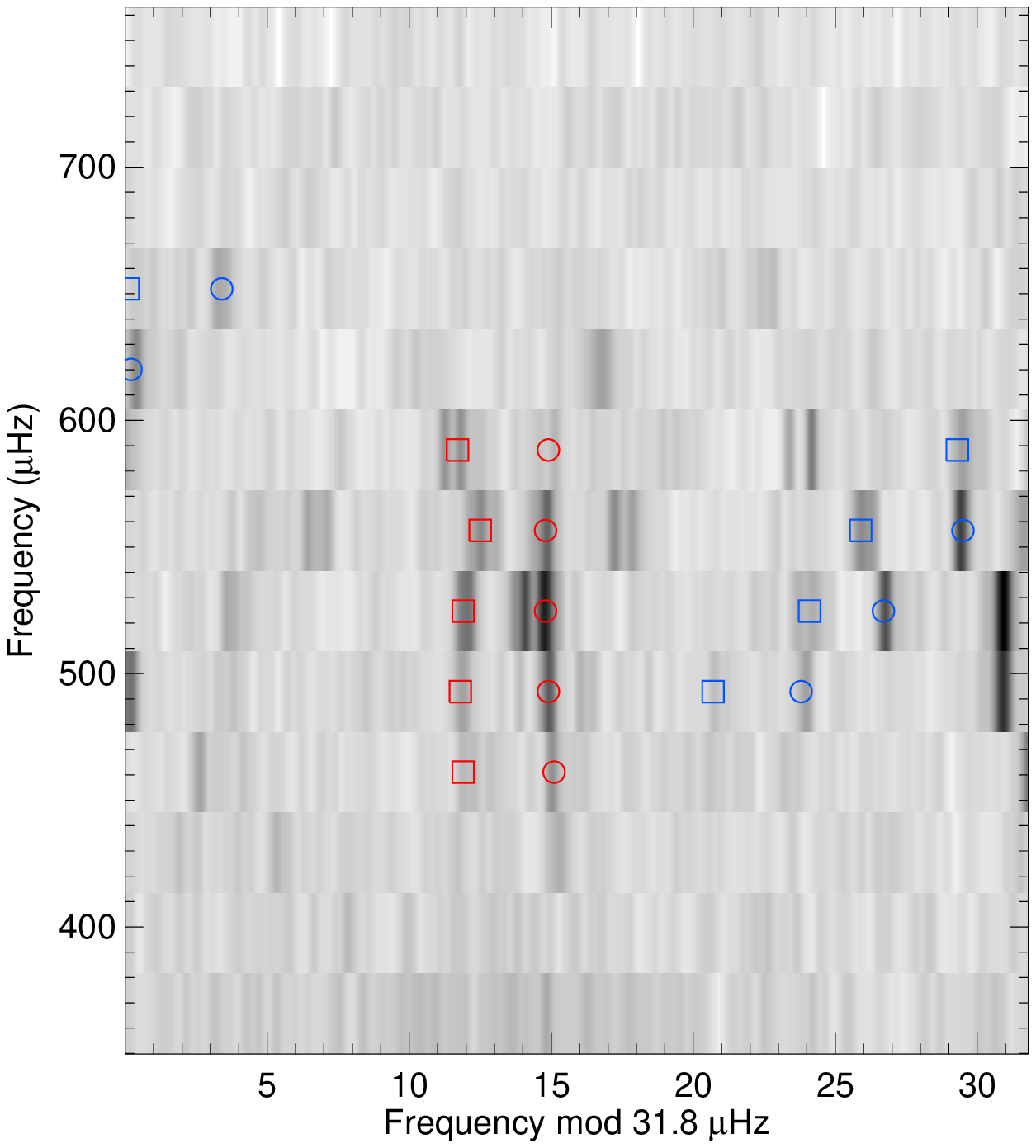}
	\includegraphics[width=7.0cm]{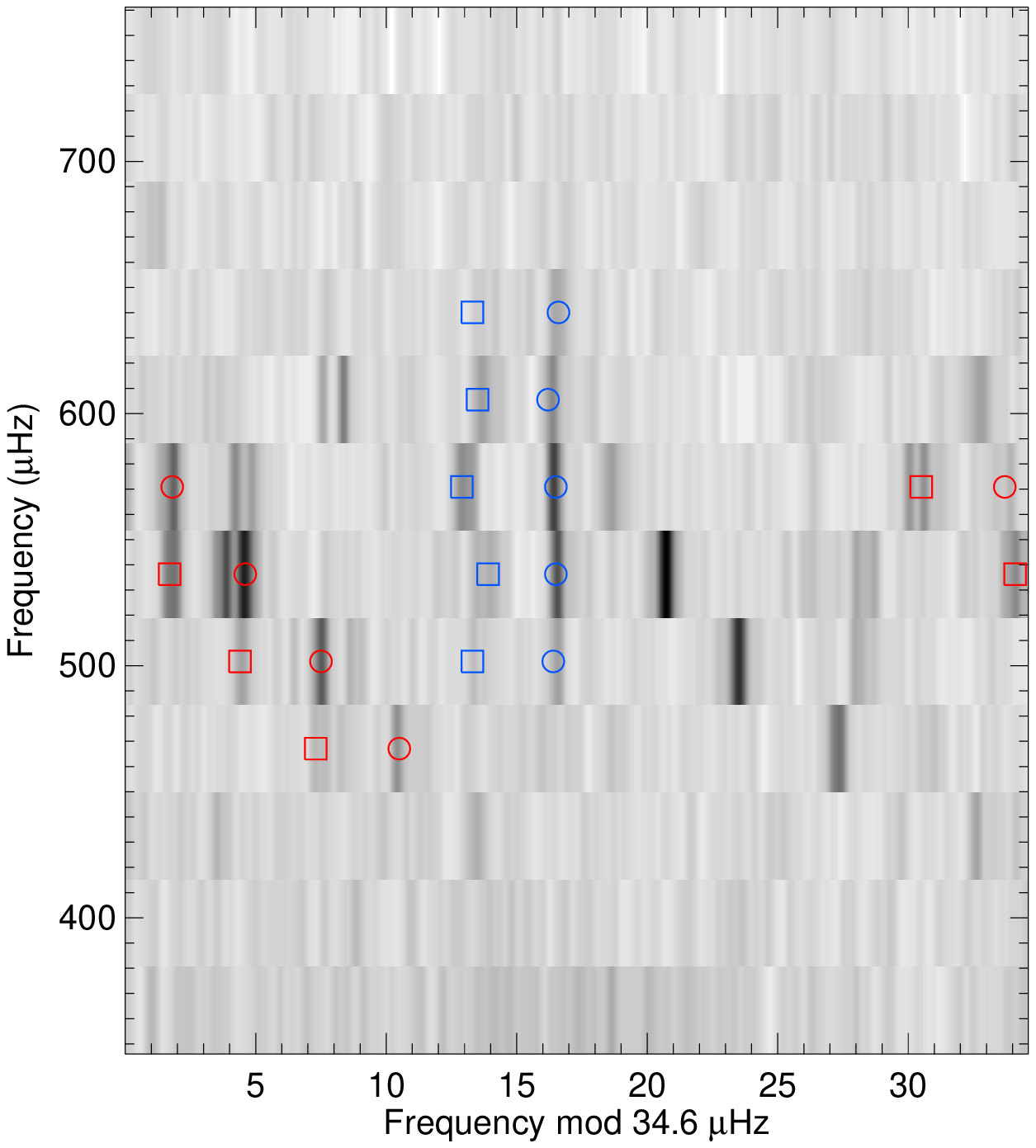}
    \caption{\'{E}chelle diagrams of star A (left) and star B (right). The circular and square open symbols represent identified $l=0$ and $l=2$ modes. Red and blue colors denote modes for star A and star B, respectively. The background grey scale represents power density. Higher power density are shown darker. The unmarked peaks are $l=1$ mixed modes.}
    \label{fig:echelle}
\end{figure*}

To model the power spectrum, we used high-DImensional And multi-MOdal NesteD Sampling code (\texttt{DIAMONDS}; \citealt{2014A&A...571A..71C}). The \texttt{DIAMONDS} code utilizes Bayes' theorem:
\begin{equation}
\label{bayes}
p(\theta\mid{D,M})=\frac{{\mathcal L}(\theta\mid{D,M})\pi(\theta\mid{M})}{p(D\mid{M})},
\end{equation}
where $\theta=(\theta_1,\theta_2,...,\theta_k)$, ${\mathcal L}(\theta\mid{D,M})$, $\pi(\theta\mid{M})$, $p(\theta\mid{D,M})$ are the parameter vector, likelihood function for a given model $M$ and dataset $D$, prior probability density function, and the posterior probability density function, respectively. \texttt{DIAMONDS} uses a sampling algorithm, Nested Sampling Monte Carlo (NSMC), to tackle the high-dimensional problem.

We simultaneously fitted the two stars to separate their overlapping oscillations with the following steps. First, we modelled the power spectrum, $P(\nu)$, with:
\begin{equation}
P(\nu)=W+R\left(\nu\right)\left\{\sum_{i=1}^{k}\frac{2\sqrt{2}}{\pi}\frac{ a_i^2/b_i}{1+(\nu/b_i)^{4}}+H_0{\rm exp}\left[-\frac{(\nu-\nu_{\rm max})^2}{2\sigma^2}\right]\right\},
\end{equation}
similar to the background models presented in \cite{2014A&A...570A..41K} and \cite{2015A&A...579A..83C}. The right-hand side of the equation comprises a flat white noise $W$, a sum of Harvey power profiles \citep{1985ESASP.235..199H} with parameters $(a_i,b_i)$, and a Gaussian envelope with $(H_0,{\nu_{\rm max}},\sigma)$. The Harvey profile models stellar background caused by granulation. In the case of KIC 7107778, the background was well-fitted using three Harvey profiles, i.e. $k=3$. The Gaussian envelope denotes the range of oscillation for two overlapping stars. All these components except the noise are modulated by the response function,
\begin{equation}
R\left(\nu\right)={\rm sinc}^2\left(\frac{\pi\nu}{2\nu_{\rm Nyq}}\right),
\end{equation}
where $\nu_{\rm Nyq}=8496.36$ $\mu$Hz denotes the Nyquist frequency. The total number of free variables is $1+2\times3+3=10$. The results are shown in Table~\ref{tab:bgpara}. As argued by \cite{2014A&A...570A..41K}, the granulation frequencies $b_i$ should scale with $\nu_{\rm max}$. They provided empirical relations $b_2 = 0.317\left(\nu_{\rm max}/ \mu{\rm Hz} \right)^{0.970} \mu{\rm Hz}$ and $b_3 = 0.948\left(\nu_{\rm max} /\mu{\rm Hz}\right)^{0.992} \mu{\rm Hz}$, which result in $b_2 = 148.87$ $\mu$Hz and $b_3 = 511.87$ $\mu$Hz in our case. $b_2$ is similar to our fit but not $b_3$. This should be expected because the power spectrum is a superposition of the twins. $b_1$ should be treated carefully because the light curve was processed by a high-pass filter, which affected the low-frequency spectrum.

\begin{table}
	\centering
	\caption{Granulation background parameters.}
	\label{tab:bgpara}
	\begin{tabular}{lcc}
		\hline
Parameter & Value & 68.3\% credible \\
\hline
$W$ $[{\rm ppm}^2/\mu{\rm Hz}]$ & 11.927 & 0.100 \\
$a_1$ $[\rm ppm]$ & 58.874 & 8.314 \\
$b_1$ $[\mu{\rm Hz}]$& 5.203 & 1.460 \\
$a_2$ $[\rm ppm]$ & 67.167 & 8.705 \\
$b_2$ $[\mu{\rm Hz}]$& 149.144 & 32.207 \\
$a_3$ $[\rm ppm]$ & 76.134 & 7.947 \\
$b_3$ $[\mu{\rm Hz}]$& 400.646 & 64.318 \\
$H_0$ $[{\rm ppm}^2/\mu{\rm Hz}]$& 17.977 & 2.852 \\
$\nu_{\rm max}$ $[\mu{\rm Hz}]$& 568.051 & 7.935 \\
$\sigma$ $[\mu{\rm Hz}]$& 55.811 & 8.690 \\
\hline
	\end{tabular}
\end{table}

Then, each mode was fitted with a Lorentzian profile with three free parameters: frequency centroid $\nu_0$, amplitude $A$ and linewidth $\Gamma$ built on the background:
\begin{equation}
L(\nu)=R\left(\nu\right)\left[\frac{A^2/\pi\Gamma}{1+4\left(\nu-\nu_0\right)^2/\Gamma^2}\right].
\end{equation}
 The power spectrum was fitted with a sum of 32 Lorentzian profiles. The total number of free variables is $3\times32=96$. It is possible that the star rotates and lifts degeneracy of $m$-degree of the modes \citep{2003ApJ...589.1009G}. Thus we performed a hypothesis test with Bayesian evidence denoted by $p(D\mid{M})$ in Equation~\ref{bayes}. The Bayesian evidence balances the goodness of fit and the need to fit, as it can be written as a product of the maximum of the likelihood and an Occam factor \citep{2014arXiv1411.3013K}. We adopted a splitting model of $l=1$ modes
\begin{equation}
L(\nu)=R\left(\nu\right)\left[\sum_{m=-1}^{1}\frac{\xi_m A^2/\pi\Gamma}{1+4\left(\nu-\nu_0-m\delta\nu_{\rm split}\right)^2/\Gamma^2}\right]
\end{equation}
where $i$ is the inclination angle, $\xi_{-1} = \xi_{1} = 0.5\sin^2{i}$, $\xi_{0} = \cos^2{i}$ and $\delta\nu_{\rm split}$ measures the extent of splitting. We fitted the splitting model to four $l=1$ modes (mode frequency 477, 507, 539 and 620 $\mu$Hz) individually. The four modes were selected because they do not have another mode in $3$ $\mu$Hz frequency range. We defined detection probability as $p = p(D\mid{M_B})/(p(D\mid{M_A})+p(D\mid{M_B}))$ \citep{2014A&A...571A..71C,jeffreys1961theory}, where subscripts A and B refer to the non-splitting model and the splitting model. The detection probabilities are small: $10^{-1}$, $10^{-14}$, $10^{-9}$ and $10^{-2}$ respectively. We also fitted the four modes with a common inclination and the detection probability is $10^{-50}$. This indicates the non-splitting model gives a better depiction of the data. Hence, detecting the inclination angle $i$ through rotation was not considered.

Notice that only $l=0$ and $l=2$ modes can be identified straightforwardly and allocated to one of the two stars, since they are regularly spaced. The $l=1$ mixed modes are strongly bumped and no clear patterns could be followed, so we used stellar models to help the identification, as discussed in Section~\ref{sec:model}. We also measured the frequency of maximum power, $\nu_{\rm max}$, by fitting a Gaussian profile in power density to the $l=0$ mode peaks. The results are $\nu_{\rm max,A}=523\pm16$ $\mu$Hz and $\nu_{\rm max,B}=570\pm18$ $\mu$Hz, where the uncertainties are half of $\langle\Delta\nu\rangle$.

Fig.~\ref{fig:echelle} shows the \'{e}chelle diagram of both stars. The circular and square open symbols represent identified $l=0$ and $l=2$ modes. Red and blue colors denote star A and star B, respectively. The mode parameters are shown in Tables~\ref{tab:modeA} and~\ref{tab:modeB}. The $l=1$ modes in the tables are further discussed in Section~\ref{subsec:dual}.

\section{Asteroseismic Analysis}
\label{sec:model}

As discussed in section~\ref{sec:obs}, KIC 7107778 has a long orbital period and no evidence of eclipses from which stellar fundamental parameters could be estimated. Therefore, estimating them through stellar models with asteroseismology is necessary.

\subsection{Stellar Models}
\label{subsec:models}

We constructed stellar models using Modules for Experiments in Stellar Astrophysics (\texttt{MESA}; \citealt{2011ApJS..192....3P}; \citealt{2013ApJS..208....4P}; \citealt{2015ApJS..220...15P}). \cite{2011ApJS..192....3P} have discussed the input physics of \texttt{MESA}. Here we list them for completeness. The equation of state was delivered by OPAL EOS tables \citep{2002ApJ...576.1064R}, SVCH tables \citep{1995ApJS...99..713S} at low temperatures and densities, and HELM \citep{2000ApJS..126..501T} and PC \citep{2010CoPP...50...82P} tables under other circumstances. Opacities were taken from \cite{1996ApJ...464..943I} at high temperature and \cite{2005ApJ...623..585F} at low temperature. Nuclear reaction rates were based on NACRE \citep{1999NuPhA.656....3A} and CF88 \citep{1988ADNDT..40..283C} when NACRE was unavailable. The convection was implemented with Mixing Length Theory (MLT) illustrated in \cite{1968pss..book.....C}. The mixing length parameter $\alpha_{\rm MLT}=1.917$ was employed according to the \texttt{MESA} standard solar model \citep{2011ApJS..192....3P}. Overshoot mixing was set according to \cite{2000A&A...360..952H}, with overshooting parameter $f_{\rm ov}=0.016$. The initial helium abundance was estimated through
\begin{equation}
Y_{\rm ini}=Y_0+\frac{\Delta Y}{\Delta Z}\cdot{Z_{\rm ini}}
\end{equation}
where $Y_0=0.249$ \citep{2016A&A...594A..13P} and ${\Delta{Y}}/{\Delta{Z}}=1.33$, calculated using the initial abundances of helium and heavy elements of the calibrated solar model \citep{2011ApJS..192....3P}. The relation of metallicity and element abundance ratio we adopted here was
\begin{equation}
\rm{[Fe/H]}=\log(Z/X)-{\log}{(Z/X)_{\astrosun}},
\end{equation}
where the solar value is $(Z/X)_{\astrosun}=0.02293$ \citep{1998SSRv...85..161G}.

The varying input parameters for grid modelling are mass $M$ and metallicity [Fe/H]. They were set as follows. First, the mass was set according to asteroseismic scaling relations. The mean large separation $\Delta\nu$ and frequency of maximum power $\nu_{\rm max}$ are related to mean density $\rho$, surface gravity $g$ and effective temperature $T_{\rm{eff}}$: $\Delta\nu\propto\sqrt{\rho}$, $\nu_{\rm{\rm max}}\propto{g/\sqrt{T_{\rm{eff}}}}$ \citep{1995A&A...293...87K}, i.e.
\begin{equation}
\frac{\Delta\nu}{\Delta\nu_{\astrosun}}\approx\left(\frac{M}{M_{\astrosun}}\right)^{1/2}\left(\frac{R}{R_{\astrosun}}\right)^{-3/2},
\label{eq:dnu}
\end{equation}
\begin{equation}
\frac{\nu_{\rm max}}{\nu_{\rm max,\astrosun}}\approx\left(\frac{M}{M_{\astrosun}}\right)\left(\frac{R}{R_{\astrosun}}\right)^{-2}\left(\frac{T_{\rm{eff}}}{T_{\rm{eff},\astrosun}}\right)^{-1/2},
\label{eq:numax}
\end{equation}
where $\Delta\nu_{\astrosun}=135.1$ $\mu$Hz, $\nu_{\rm max,\astrosun}=3050$ $\mu$Hz \citep{2013ARA&A..51..353C} and $T_{\rm{eff}}=5777$ K.
Thus, the mass prescription can be deduced from equation~\ref{eq:dnu} and~\ref{eq:numax}:
\begin{equation}
\frac{M}{M_{\astrosun}}\approx\left(\frac{\Delta\nu}{\Delta\nu_{\astrosun}}\right)^{-4}\left(\frac{\nu_{\rm{max}}}{\nu_{\rm{max},\astrosun}}\right)^{3}\left(\frac{T_{\rm{eff}}}{T_{\rm{eff},\astrosun}}\right)^{3/2}.
\end{equation}
Considering the uncertainties, the mass range of the grid should at least cover 1.34 to 1.60 $M_{\astrosun}$. Second, we adopted metallicity [Fe/H]=0.01, 0.11 and 0.21 according to spectral observations from LAMOST. Table~\ref{tab:parameter} summarizes the input parameters of our grid modelling.

\begin{table}
	\centering
	\caption{Input parameters of grid modelling.}
	\label{tab:parameter}
	\begin{tabular}{lccc}
		\hline
		  & Value & Step size \\
		\hline
		$M$ $[M_{\astrosun}]$ & 1.34$\sim$1.60 &  0.1 \\
		{[Fe/H]} $[\rm dex]$& 0.01$\sim$0.21  & 0.1 \\
		$\alpha_{MLT}$ & 1.917 &  fixed\\
		$f_{ov}$ & 0.016 &   fixed\\
		\hline
	\end{tabular}
\end{table}

\subsection{Modelling the $l=0$ and $l=2$ modes}
\label{subsec:constraint}

We calculated oscillation frequencies for models that met the requirements of $\Delta\nu$ and $\nu_{\rm{max}}$ with \texttt{GYRE} \citep{2013MNRAS.435.3406T}, which solves the adiabatic pulsation equations with stellar structure data.

The calculated frequencies deviated from the observations due to the improper simulation of the stellar surface. Therefore, we followed the method described by \cite{2014A&A...568A.123B} to correct them. The correction formula we adopted here was:
\begin{equation}
\delta\nu=(a_{-1}\nu^{-1}+a_3\nu^3)/I,
\end{equation}
where $a_{-1}$ and $a_3$ are coefficients determined through least-squares fit, and $I$ is the mode inertia.

We used $\chi^2=\Sigma(\nu_{\rm obs}-\nu_{\rm mod})^2/\sigma^2$ as the quality of the fit. Frequencies used in the $\chi^2$ calculation were all $l=0$ and $l=2$ modes. The smaller the $\chi^2$ of the model, the better the match. We took the best 10\% of the models, as measured by $\chi^2$, for further considerations. The choice of this criterion was a trade-off. Small values could be biased by fluctuations in limited samples. The problem was tackled by sorting samples according to the values of $\chi^2$, performing difference between two adjacent quantities. We found the trend around 10\% went smoothly, which ruled out the fluctuation effect. Including more models would make the selection less useful. This point was addressed by inspecting the value of $\chi^2$ around the 10\% cut-off and assuring that it was not too large.

\begin{figure*}
	\includegraphics[width=16.0cm]{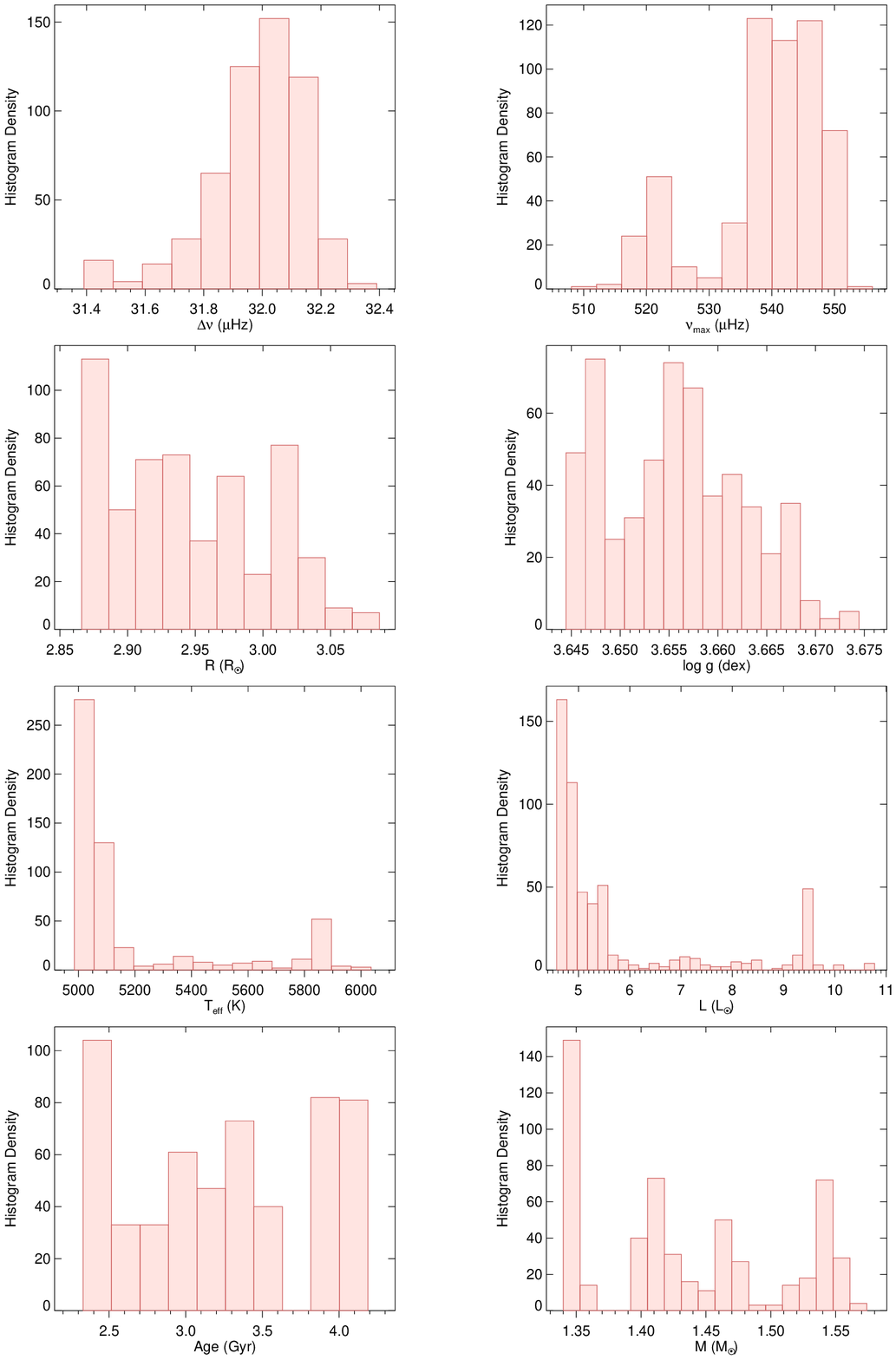}
    \caption{Parameter distributions for the lowest 10\% $\chi^2$ of star A grid models with [Fe/H]=0.11 dex.}
    \label{fig:paraA}
\end{figure*}

\begin{figure*}
	\includegraphics[width=16.0cm]{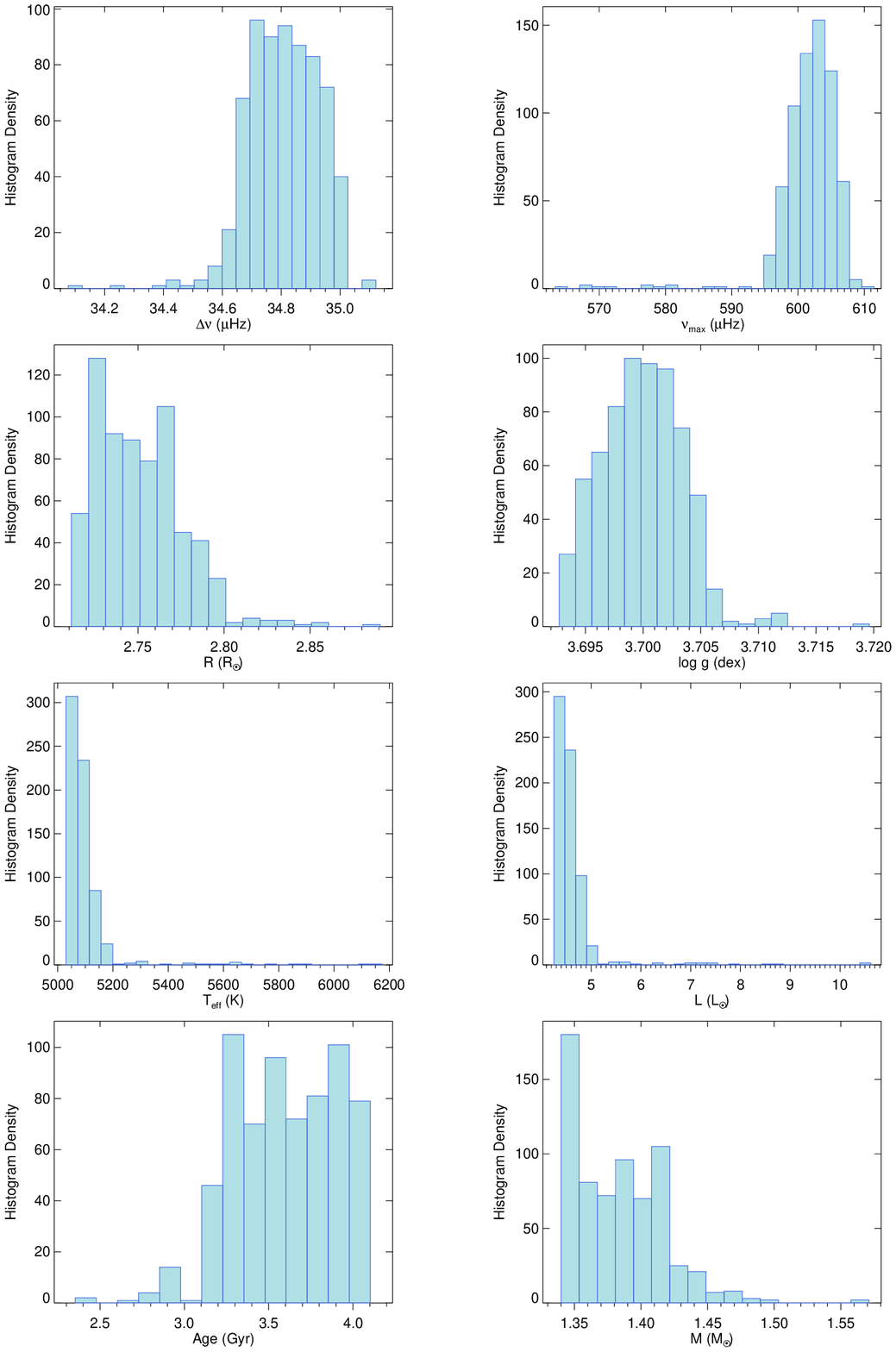}
    \caption{Parameter distributions for the lowest 10\% $\chi^2$ of star B grid models with [Fe/H]=0.11 dex.}
    \label{fig:paraB}
\end{figure*}

With the lowest 10\% $\chi^2$ models, we calculated the mean value of the stellar parameters as the centroid. Table~\ref{tab:gridmodel} lists them with standard deviations. Additionally, Fig.~\ref{fig:paraA} and~\ref{fig:paraB} display histograms of each stellar parameter, which reflect the distribution in the lowest 10\% $\chi^2$ models with [Fe/H]=0.11. Red and blue indicate star A and star B, respectively.

Considering that [Fe/H] was estimated with uncertainties, we combined stellar parameters derived based on different [Fe/H] together, as the ultimate results. Here we list mass, radius and age:  $M_A=1.42\pm0.06$ $M_{\astrosun}$, $M_B=1.39\pm0.03$ $M_{\astrosun}$, $R_A=2.93\pm0.05$ $R_{\astrosun}$, $R_B=2.76\pm0.04$ $R_{\astrosun}$, $t_A=3.32\pm0.54$ Gyr and $t_B=3.51\pm0.33$ Gyr.

\subsection{Modelling the $l=1$ mixed modes}
\label{subsec:dual}
We next searched for the best models which could also fit the frequencies of $l=1$ modes from both stars. KIC 7107778 contains two sub-giant stars. Tiny changes to the mass of the models influence oscillations greatly. Only extremely fine grids produce satisfying results, which makes the task demanding. Since our purpose was to find a pair of stellar models for both stars that fit observations best, we used the bisection method to search.

The main idea of the bisection method is summarized as follows, similar to finding solutions for equation $f(x)=0$. First, we started with two masses, $M_1$ and $M_2$, which lied on opposite sides of the best model. This choice was realized by visually inspecting oscillation frequencies on the \'{e}chelle diagram. Second, we bisected this range, i.e. calculated frequencies of $M_3=(M_1+M_2)/2$. Third, we evaluated the result of $M_3$ and determined $M_3$ and whichever of $M_1$ or $M_2$ yield the best model. Fourth, we kept bisecting the mass range until the difference became sufficiently small.

\begin{figure*}
	\includegraphics[width=7.0cm]{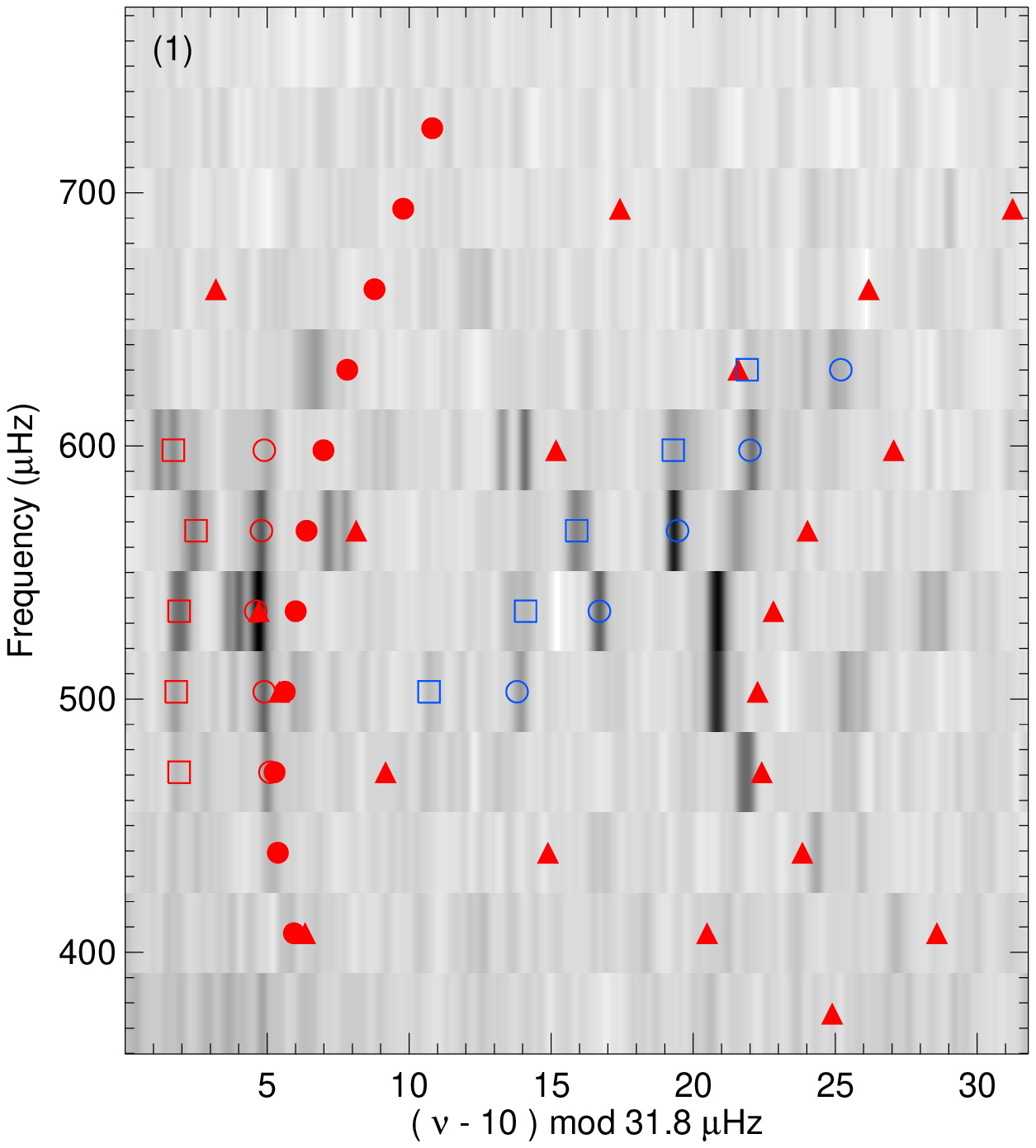}
	\includegraphics[width=7.0cm]{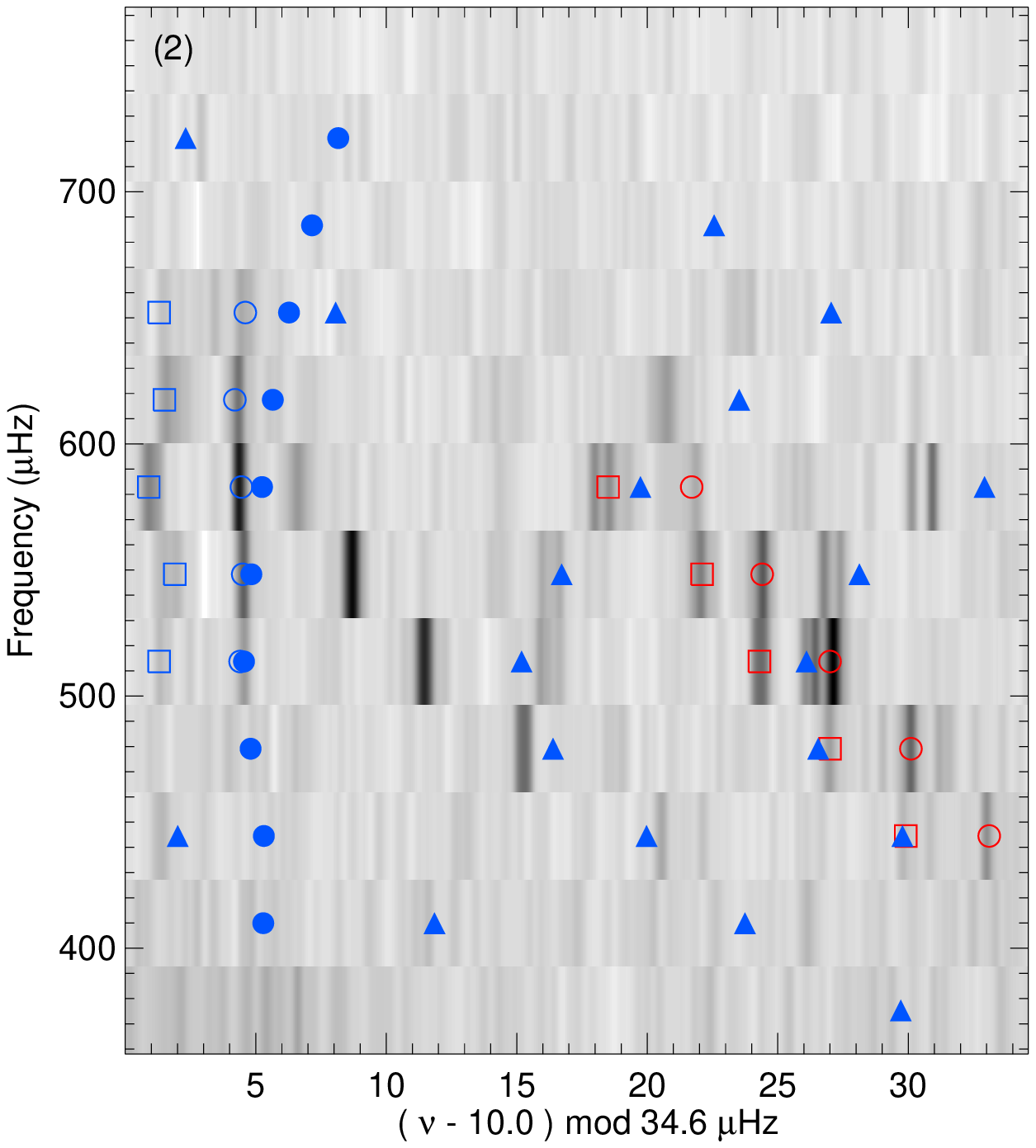}
	\includegraphics[width=7.0cm]{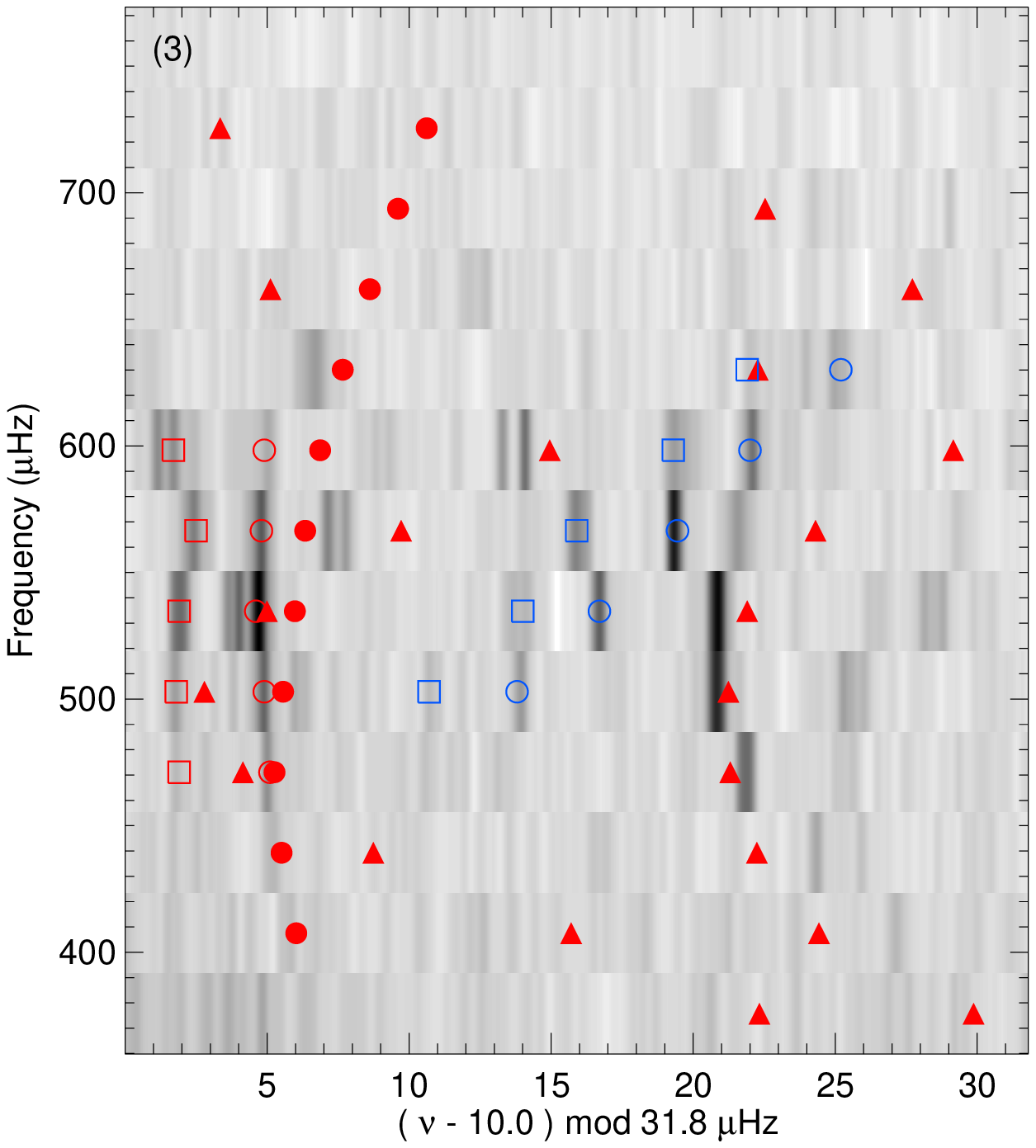}
	\includegraphics[width=7.0cm]{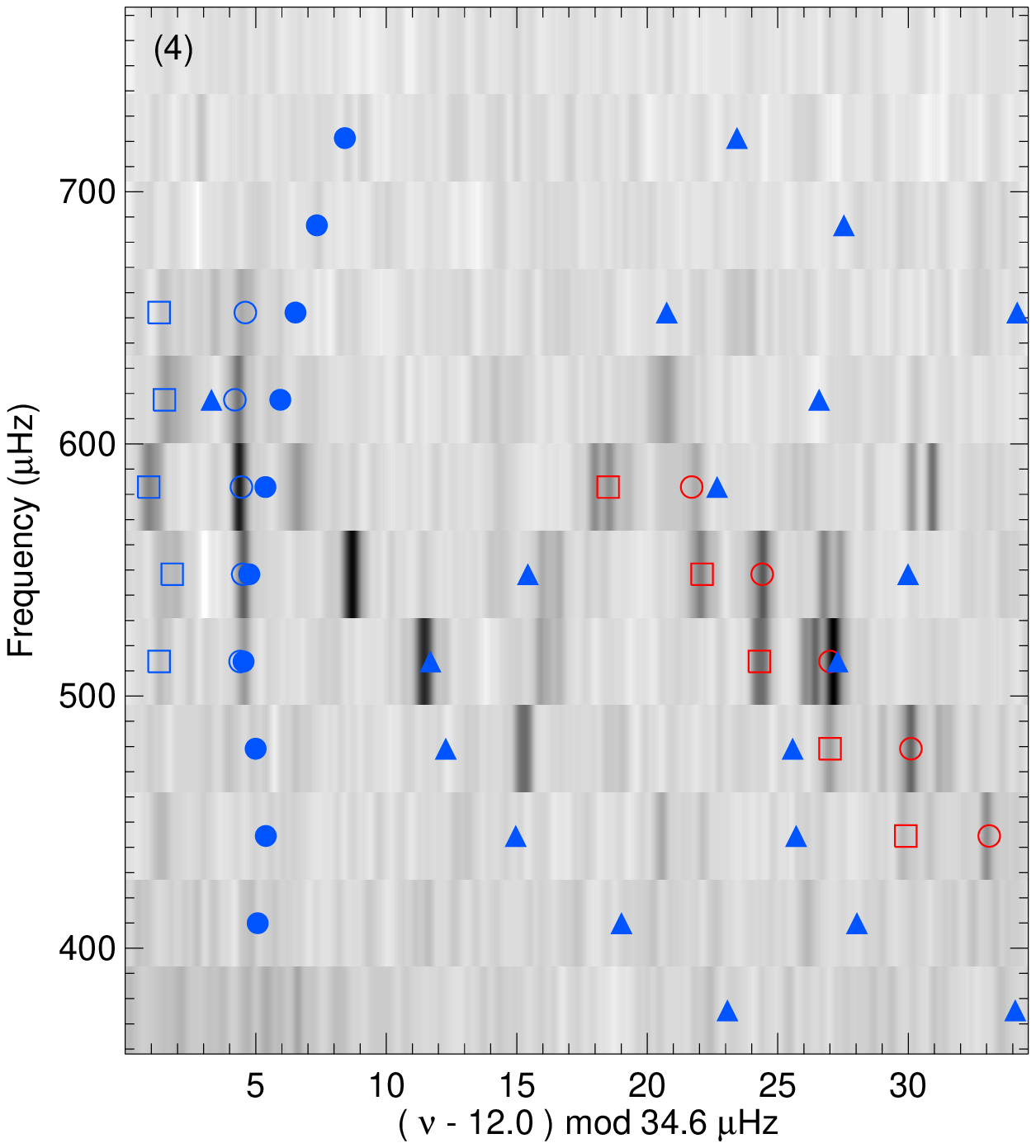}
	\includegraphics[width=7.0cm]{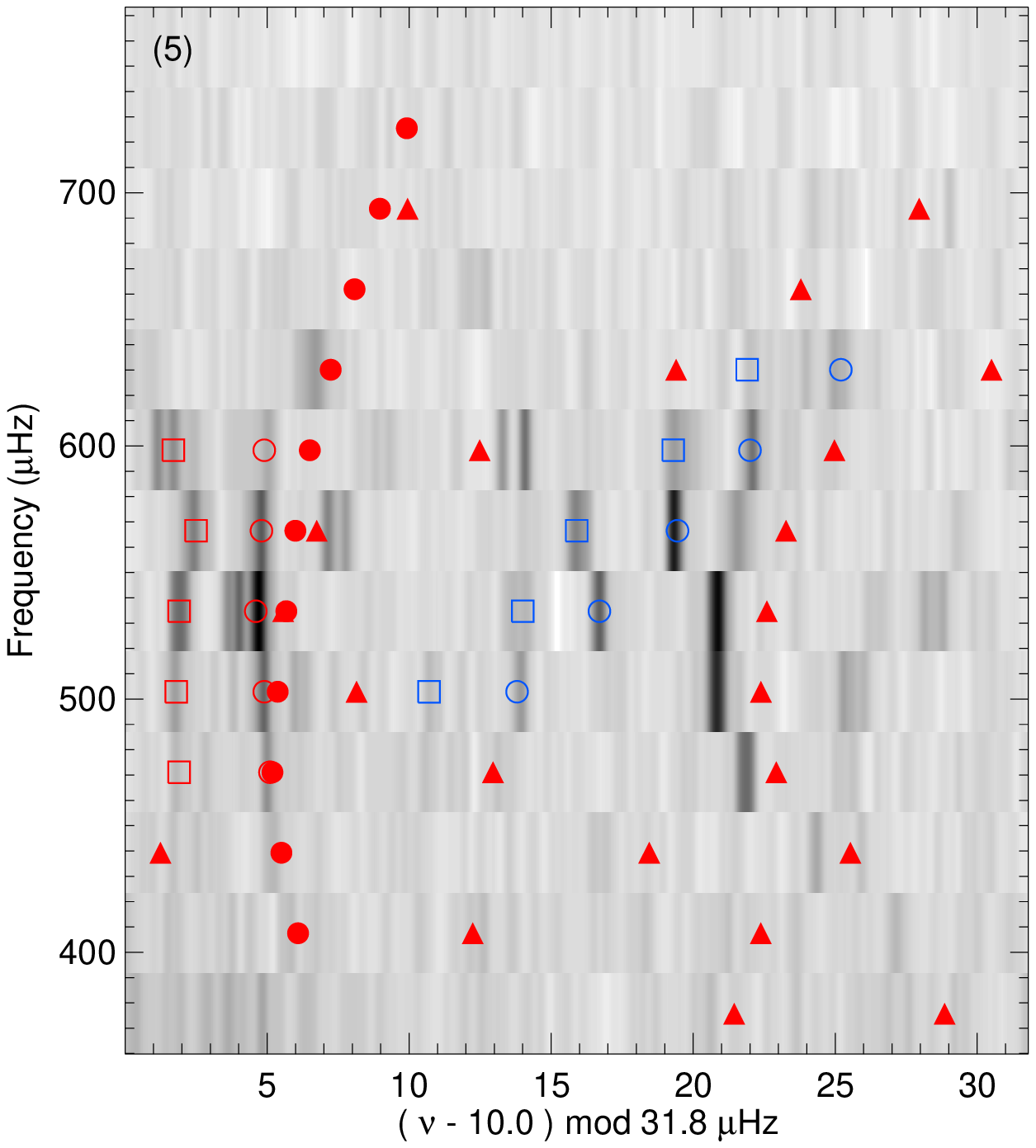}
	\includegraphics[width=7.0cm]{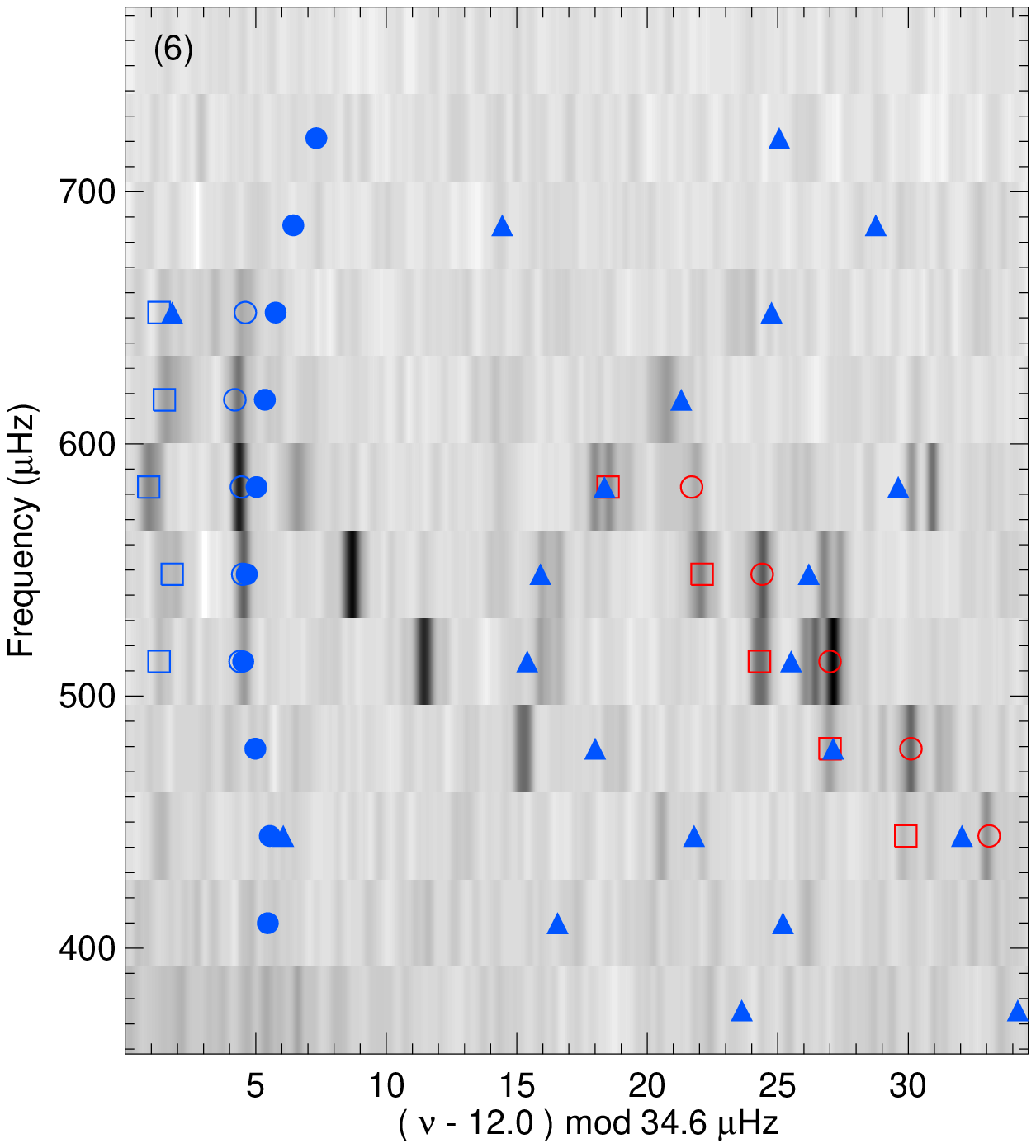}
    \caption{\'{E}chelle diagrams of KIC 7107778. Number series (1)$\sim$(6) on the upper left corner of each panel corresponds to the series of models in Table~\ref{tab:dichmodel}, whose oscillation frequencies are displayed in the corresponding \'{e}chelle. These models are without surface corrections. Open and filled symbols represent observational frequencies and theoretical ones. Red and blue represent modes of star A and star B respectively. Circles, triangles, and squares denote $l=0$, $l=1$ and $l=2$ modes.}
    \label{fig:echelleall}
\end{figure*}

We followed this search scheme under three different metallicities [Fe/H]: 0.01, 0.11 and 0.21. The maximum precision in calculation reaches 0.00001 $M_{\astrosun}$. The results revealed three combinations of best fitting models for the two stars. Although they do still deviate from observed peaks, they give a reasonable fit to most $l=1$ modes. Most model masses reach to 0.001 $M_{\astrosun}$ precision. The reason of such small precision is that the mixed modes are very sensitive to the change of internal structure, and subgiants evolve very fast. Each combination has less than 0.2 Gyr difference in age, consistent with the idea that two components formed at the same time. Table~\ref{tab:dichmodel} presents the fundamental parameters of the three model pairs, and Fig.~\ref{fig:echelleall} displays them on \'{e}chelle diagrams. Open and filled symbols represent observational and theoretical frequencies. Red and blue indicate modes of star A and star B, respectively. Circles, squares and triangles denote $l=0$, $l=1$, and $l=2$ modes.

\begin{figure}
	\includegraphics[width=9.0cm]{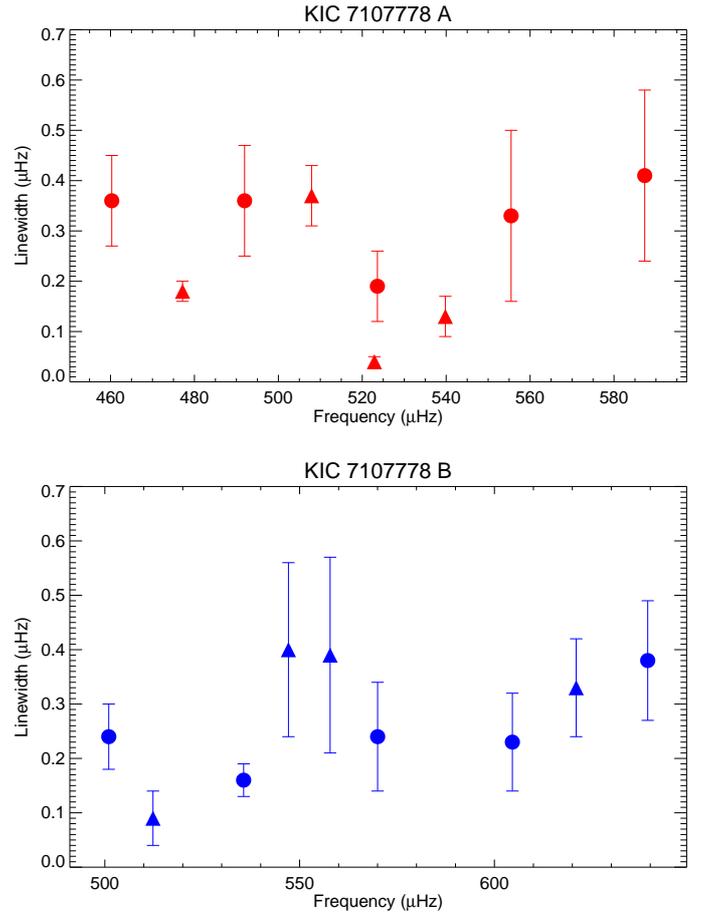}
    \caption{Mode linewidth as a function of mode frequency. The circles and triangles represent the $l=0$ and $l=1$ modes respectively. The $l=1$ modes shown here are only those which have relative certain identity (not associated with ``?'' or ``*'' in Table~\ref{tab:modeA} and \ref{tab:modeB}). The $l=2$ modes are instead not presented because each $l=2$ mode region is fitted with a Lorentzian profile, the linewidth of which is not necessarily representing the real mode lifetime.}
    \label{fig:lw}
\end{figure}

Based on theoretical models, we found that most observed peaks could be matched with $l=1$ modes. Tables~\ref{tab:modeA} and~\ref{tab:modeB} display the mode parameters in each Lorentzian profile we fitted to each peak, with associated $l$ degrees. Modes which share the same peak on the power spectrum are labeled with an asterisk mark. Some modes from the two stars stand too close and cause ambiguity; they are labeled with a question mark. Here we remind readers that this solution is not unique, considering that the models still differ from the observation. In Fig.~\ref{fig:lw}, we show the mode linewidth as a function of the mode frequency. Mixed modes are expected to have smaller linewidths compared to radial modes because they have contributions from g modes trapped in the core, resulting a longer lifetime \citep{2013ApJ...767..158B,2009A&A...506...57D}. This is strongest for the mode at 522.897 $\mu$Hz. For other $l=1$ modes, the linewidths are similar to $l=0$ modes because they are less bumped and more p-like.

\begin{figure*}
	\includegraphics[width=18.0cm]{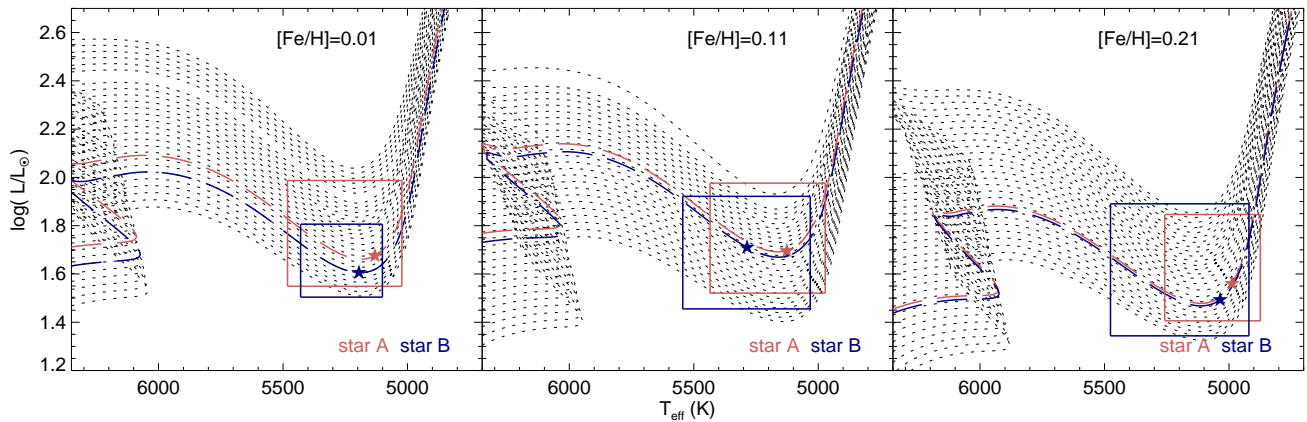}
    \caption{Hertzsprung-Russell diagram. Grid model tracks are displayed as black dotted lines. The best models, derived through bisection, are drawn with star symbols located on evolutionary tracks (dashed lines) in red for star A and in blue for star B. The solid lines indicate the standard deviation of effective temperature $T_{\rm{eff}}$ and luminosity $L$, which are shown in Table~\ref{tab:gridmodel}. The left, middle and right panel represent models for metallicity [Fe/H] 0.01, 0.11 and 0.21 respectively.}
    \label{fig:HR}
\end{figure*}

Fig.~\ref{fig:HR} is the Hertzsprung-Russel diagram where grid and additional models are displayed with dotted black lines. Specifically, the dashed lines indicate the best model tracks from the bisecting method with star A in red and star B in blue, respectively. The star symbols denote the best models. The boxes consisting of solid lines indicate the standard deviation of effective temperature $T_{\rm{eff}}$ and luminosity $L$, shown in Table~\ref{tab:gridmodel}. The left, middle and right panels represent models for metallicity [Fe/H] 0.01, 0.11 and 0.21.

\begin{table*}
	\centering
	\caption{Fundamental properties of models from grid modelling.}
	\label{tab:gridmodel}
	\begin{tabular}{ccccccccccccc}
		\hline
Star&[Fe/H]&$M$&Age&$T_{\rm{eff}}$&$L$&$R$&$\log{g}$&$\Delta\nu$&$\nu_{\rm max}$	\\
&[dex]&[$M_{\astrosun}$]&[Gyr]&[K]&[$L_{\astrosun}$]&[$R_{\astrosun}$]&[dex]&[$\mu\rm{Hz}$]&[$\mu\rm{Hz}]$	\\
		\hline
		
A&0.01&$  1.41\pm  0.06$&$ 3.178\pm 0.440$&$5216\pm 258$&$ 5.813\pm 1.508$&$ 2.925\pm 0.047$&$ 3.654\pm 0.005$&$ 32.01\pm  0.18$&$534\pm  8$&\\
B&0.01&$  1.38\pm  0.03$&$ 3.289\pm 0.263$&$5233\pm 132$&$ 5.140\pm 0.692$&$ 2.754\pm 0.025$&$ 3.699\pm 0.003$&$ 34.77\pm  0.13$&$592\pm  5$&\\
A&0.11&$  1.43\pm  0.07$&$ 3.266\pm 0.583$&$5184\pm 290$&$ 5.787\pm 1.663$&$ 2.945\pm 0.055$&$ 3.656\pm 0.007$&$ 31.98\pm  0.17$&$539\pm  9$&\\
B&0.11&$  1.38\pm  0.03$&$ 3.606\pm 0.313$&$5092\pm 105$&$ 4.594\pm 0.549$&$ 2.751\pm 0.025$&$ 3.700\pm 0.004$&$ 34.81\pm  0.12$&$601\pm  4$&\\
A&0.21&$  1.41\pm  0.07$&$ 3.643\pm 0.551$&$5037\pm 199$&$ 5.032\pm 1.160$&$ 2.926\pm 0.054$&$ 3.655\pm 0.006$&$ 32.06\pm  0.15$&$546\pm  6$&\\
B&0.21&$  1.40\pm  0.04$&$ 3.657\pm 0.317$&$5045\pm 106$&$ 4.473\pm 0.535$&$ 2.765\pm 0.028$&$ 3.701\pm 0.003$&$ 34.78\pm  0.13$&$606\pm  4$&\\
		\hline
		
	\end{tabular}
	\begin{tablenotes}
	\item \emph{Note}: the models are selected with the lowest 10\% $\chi^2$. Column 2 is the input parameter for grid modelling.
	\end{tablenotes}
\end{table*}

\begin{table*}
	\centering
	\caption{Fundamental properties of best models from bisection method.}
	\label{tab:dichmodel}
	\begin{tabular}{ccccccccccccccccc}
		\hline
NO.&Star&$M$&[Fe/H]&Age&$T_{\rm{eff}}$&$L$&$R$&$\log{g}$&$M_{H,core}$&$\Delta\nu$&$\nu_{\rm max}$	\\
{\#}&&[$M_{\astrosun}$]&[dex]&[Gyr]&[K]&[$L_{\astrosun}$]&[$R_{\astrosun}$]&[dex]&[$M_{\astrosun}$]&[$\mu\rm{Hz}$]&[$\mu\rm{Hz}$]	\\
		\hline
1&A&  1.41&0.01& 3.063&5130& 5.342& 2.929& 3.655& 0.165& 32.04&540&\\
2&B&  1.39&0.01& 3.229&5195& 4.982& 2.759& 3.699& 0.157& 34.74&594&\\
3&A&  1.48&0.11& 2.873&5120& 5.464& 2.975& 3.661& 0.170& 32.00&548&\\
4&B&  1.47&0.11& 2.929&5267& 5.478& 2.814& 3.705& 0.162& 34.64&599&\\
5&A&  1.42&0.21& 3.533&4987& 4.763& 2.927& 3.656& 0.165& 32.10&549&\\
6&B&  1.41&0.21& 3.548&5038& 4.443& 2.770& 3.703& 0.160& 34.80&608&\\
		\hline
	\end{tabular}
	\begin{tablenotes}
	\item \emph{Note}: column 3 and 4 are the input parameters for grid modelling.
	\end{tablenotes}
	
\end{table*}

\section{Conclusion}
\label{sec:con}


We applied asteroseismology to a binary target KIC 7107778, and confirmed that the two stars are in the sub-giant phase. We successfully identified their $l=0$ and $l=2$ oscillation modes and distinguished $l=1$ modes to the greatest extent. We derived stellar fundamental parameters for the two stars: $M_A=1.43^{+0.08}_{-0.08}$ $M_{\astrosun}$, $M_B=1.40^{+0.05}_{-0.06}$ $M_{\astrosun}$, $R_A=2.94^{+0.06}_{-0.06}$ $R_{\astrosun}$, $R_B=2.77^{+0.04}_{-0.04}$ $R_{\astrosun}$, $t_A=3.19^{+0.60}_{-0.64}$ Gyr and $t_B=3.26^{+0.40}_{-0.40}$ Gyr. All the evidence suggests that they formed at the same time and possess nearly equal masses.

The results yield the similarity of masses for two stars, and the best models, derived through bisection modelling, determined the mass difference as 1.42\%, 0.68\% and 0.70\%, from which we conclude it is $\sim$1\%. The H-R diagram verifies this, with extremely close tracks.

Table~\ref{tab:gridmodel} indicates the sensitivity of metallicity to age, and the two stars' ages are equal within the error. We could conclude that they formed at the same time, as is expected by binary formation that they originate from the same protostellar cloud.


The KIC 7107778 system contains two extremely similar components, with fully overlapping power spectra. This is one of the few identical twin systems to be found, proving the full potential of asteroseismology.

\section*{Acknowledgements}
We acknowledge the data from the \emph{Kepler} Discovery mission, whose funding is provided by NASA's Science Mission Directorate. This work has also made use of data from the European Space Agency (ESA) mission {\it Gaia}\footnote{\url{http://www.cosmos.esa.int/gaia}}, processed by the {\it Gaia} Data Processing and Analysis Consortium (DPAC\footnote{\url{http://www.cosmos.esa.int/web/gaia/dpac/consortium}}). Funding for the DPAC has been provided by national institutions, in particular the institutions participating in the {\it Gaia} Multilateral Agreement. S.B. and Y.L. acknowledges the Joint Research Fund in Astronomy (U1631236) under cooperative agreement between the National Natural Science Foundation of China (NSFC) and Chinese Academy of Sciences (CAS), NSFC 11273007 and 10933002, and the Fundamental Research Funds for the Central Universities. E.C. has received fundings from the European Union's Horizon 2020 research and innovation programme under the Marie Sklodowska-Curie grant agreement n$^\circ$ 664931.




\bibliographystyle{mnras}
\bibliography{binary} 








\bsp	

\begin{table*}
	\centering
	\caption{Mode parameters of KIC 7107778 A.}
	\label{tab:modeA}
	\begin{tabular}{ccccccccc}
		\hline
$l$	& Frequency &68.3\% credible&	Amplitude &68.3\% credible&	Linewidth&68.3\% credible& 		&	\\
	& [$\mu$Hz] & [$\mu$Hz]		& [ppm]&[ppm]&[$\mu$Hz] & [$\mu$Hz]	&		&	\\
		\hline
0&460.294&$-  0.022/+  0.021$&  15.66&$-   0.80/+   1.19$&   0.36&$-   0.09/+   0.10$& & \\
0&491.947&$-  0.013/+  0.014$&  24.17&$-   1.29/+   1.66$&   0.36&$-   0.10/+   0.13$& & \\
0&523.603&$-  0.012/+  0.018$&  25.84&$-   1.85/+   1.35$&   0.19&$-   0.07/+   0.06$& & \\
0&555.473&$-  0.133/+  0.090$&  22.65&$-   5.44/+   2.56$&   0.33&$-   0.16/+   0.18$& & \\
0&587.314&$-  0.420/+  1.462$&   9.65&$-   2.24/+   3.40$&   0.41&$-   0.17/+   0.18$& & \\
		\hline
1&477.158&$-  0.031/+  0.042$&  15.47&$-   0.18/+   0.25$&   0.18&$-   0.02/+   0.02$& & \\
1&488.837&$-  0.022/+  0.019$&  10.70&$-   0.48/+   0.64$&   0.26&$-   0.09/+   0.07$&*& \\
1&507.928&$-  0.014/+  0.013$&  25.92&$-   0.83/+   0.99$&   0.37&$-   0.06/+   0.06$& & \\
1&522.897&$-  0.002/+  0.002$&  21.55&$-   0.75/+   1.00$&   0.04&$-   0.01/+   0.01$& & \\
1&539.794&$-  0.011/+  0.008$&  32.01&$-   1.36/+   2.02$&   0.13&$-   0.04/+   0.05$& & \\
1&558.446&$-  0.172/+  0.072$&   9.08&$-   3.06/+   3.49$&   0.06&$-   0.02/+   0.04$& &?\\
1&572.522&$-  0.827/+  0.844$&   9.03&$-   2.80/+   4.84$&   0.24&$-   0.13/+   0.09$& &?\\
1&595.812&$-  0.040/+  0.042$&  12.35&$-   1.44/+   2.36$&   0.17&$-   0.11/+   0.08$& &?\\
		\hline
2&457.106&$-  0.046/+  0.041$&  11.10&$-   0.49/+   0.40$&   0.41&$-   0.07/+   0.07$& & \\
2&488.837&$-  0.022/+  0.019$&  10.70&$-   0.48/+   0.64$&   0.26&$-   0.09/+   0.07$&*& \\
2&520.851&$-  0.020/+  0.024$&  19.66&$-   0.31/+   0.64$&   0.19&$-   0.05/+   0.04$&*& \\
2&553.122&$-  0.321/+  0.354$&  16.27&$-   2.43/+   3.94$&   0.43&$-   0.16/+   0.14$& & \\
2&584.175&$-  0.393/+  0.370$&  15.78&$-   2.33/+   2.80$&   0.38&$-   0.13/+   0.14$&*& \\
		\hline
	\end{tabular}
	\begin{tablenotes}
	\item \emph{Note}: Modes which share the same peak on the power spectrum are labeled with ``*'' marks. Modes from two stars standing too close and causing ambiguity are denoted with ``?'' marks.
	\end{tablenotes}
\end{table*}

\begin{table*}
	\centering
	\caption{Mode parameters of KIC 7107778 B.}
	\label{tab:modeB}
	\begin{tabular}{ccccccccc}
		\hline
$l$	& Frequency &68.3\% credible&	Amplitude &68.3\% credible&	Linewidth&68.3\% credible& 		&	\\
	& [$\mu$Hz]& [$\mu$Hz]		& [ppm]&[ppm]&[$\mu$Hz] & [$\mu$Hz]	&		&	\\
		\hline
0&501.006&$-  0.019/+  0.020$&  14.08&$-   0.55/+   1.06$&   0.24&$-   0.07/+   0.05$& & \\
0&535.615&$-  0.016/+  0.017$&  19.47&$-   0.66/+   0.94$&   0.16&$-   0.03/+   0.04$& & \\
0&570.025&$-  0.146/+  0.423$&  27.22&$-   6.39/+   8.01$&   0.24&$-   0.11/+   0.10$& & \\
0&604.609&$-  0.026/+  0.022$&  15.92&$-   2.44/+   1.51$&   0.23&$-   0.08/+   0.11$& & \\
0&639.344&$-  0.076/+  0.068$&  13.11&$-   0.74/+   0.99$&   0.38&$-   0.11/+   0.13$& & \\
		\hline
1&512.368&$-  0.008/+  0.012$&  11.26&$-   0.93/+   1.19$&   0.09&$-   0.06/+   0.04$& & \\
1&520.851&$-  0.020/+  0.024$&  19.66&$-   0.31/+   0.64$&   0.19&$-   0.05/+   0.04$&*& \\
1&547.122&$-  0.735/+  2.480$&  11.15&$-   2.42/+   4.08$&   0.40&$-   0.16/+   0.17$& & \\
1&557.829&$-  0.177/+  0.168$&  15.37&$-   5.30/+   5.61$&   0.39&$-   0.19/+   0.17$& &?\\
1&584.175&$-  0.393/+  0.370$&  15.78&$-   2.33/+   2.80$&   0.38&$-   0.13/+   0.14$&*& \\
1&596.595&$-  0.001/+  0.001$&  18.12&$-   1.00/+   1.37$&   0.02&$-   0.00/+   0.01$& & \\
1&620.979&$-  0.049/+  0.040$&  16.97&$-   1.01/+   1.17$&   0.33&$-   0.07/+   0.10$& & \\
		\hline
2&497.849&$-  0.096/+  0.091$&  10.21&$-   0.74/+   0.89$&   0.38&$-   0.11/+   0.12$& & \\
2&532.691&$-  0.342/+  0.315$&  11.42&$-   1.28/+   2.08$&   0.41&$-   0.14/+   0.14$& & \\
2&566.534&$-  0.731/+  0.830$&  10.72&$-   2.78/+   7.32$&   0.44&$-   0.17/+   0.30$& & \\
2&601.946&$-  0.106/+  0.109$&  14.16&$-   0.93/+   1.22$&   0.36&$-   0.10/+   0.10$& & \\
2&636.541&$-  0.380/+  0.493$&   9.04&$-   1.04/+   1.20$&   0.44&$-   0.14/+   0.13$& & \\
		\hline
	\end{tabular}
	\begin{tablenotes}
	\item \emph{Note}: Modes which share the same peak on the power spectrum are labeled with ``*'' marks. Modes from two stars standing too close and causing ambiguity are denoted with ``?'' marks.
	\end{tablenotes}
\end{table*}

\clearpage


\label{lastpage}
\end{document}